\documentclass[%
 aip,
 amsmath,amssymb,
 reprint,%
]{revtex4-1}

\usepackage{graphicx}
\usepackage{dcolumn}
\usepackage{bm}

\usepackage[utf8]{inputenc}
\usepackage[T1]{fontenc}
\usepackage{mathptmx}
\usepackage{etoolbox}

\makeatletter
\def\@email#1#2{%
 \endgroup
 \patchcmd{\titleblock@produce}
  {\frontmatter@RRAPformat}
  {\frontmatter@RRAPformat{\produce@RRAP{*#1\href{mailto:#2}{#2}}}\frontmatter@RRAPformat}
  {}{}
}%
\makeatother
\begin{document}
\title{Minimally-diffracting quartz for ultra-low temperature surface acoustic wave resonators} 



\author{A. L. Emser}
    \email[]{alec.emser@colorado.edu}
\author{B. C. Rose}
\author{L. R. Sletten}
\author{P. Aramburu Sanchez}
\author{K. W. Lehnert}

\affiliation{JILA, National Institute of Standards and Technology and the University of Colorado, Boulder, Colorado 80309, USA}
\affiliation{Department of Physics, University of Colorado, Boulder, Colorado 80309, USA}


\date{\today}

\begin{abstract}
We simulate and experimentally demonstrate the existence of an orientation of quartz which minimizes diffraction losses in surface acoustic wave (SAW) resonators at ultra-low temperatures. The orientation is optimized for applications to quantum technologies which benefit from high mechanical quality factors, strong electromechanical coupling, and narrow acoustic apertures. We fabricate narrow aperture SAW resonators on this substrate and measure internal quality factors greater than 100,000 at mK temperatures.

\end{abstract}
\maketitle 

Acoustic systems are a promising resource which offer quantum technologies a favorable combination of compact footprints\cite{morgan2010surface,chu2020perspective}, excellent coherence times\cite{kharel2018ultra,maccabe2020nano,bereyhi2022perimeter}, and the ability to connect disparate quantum systems\cite{maity2020coherent, mirhosseini2020superconducting, delaney2022superconducting, decrescent2022tightly}. The emergent field of circuit quantum acoustodynamics (cQAD) has leveraged these advantages to create hybrid platforms which are capable of exploring fundamental quantum physics \cite{satzinger2018quantum, chu2018creation, sletten2019resolving, wollack2022quantum, Lupke2022parity} and offer the potential for quantum computation with acoustic processors \cite{chamberland2022building,arrangoiz2019QAP,Hann2019}. In particular, cQAD experiments utilizing surface acoustic wave (SAW) resonators have demonstrated increasingly sophisticated quantum control over phonons, including phonon number counting \cite{sletten2019resolving}, phonon-mediated qubit-qubit entanglement\cite{dumur2021quantum}, and multipartite phonon entanglement \cite{andersson2022squeezing}. 

\begin{figure} 
\centering
\includegraphics[width=\columnwidth]{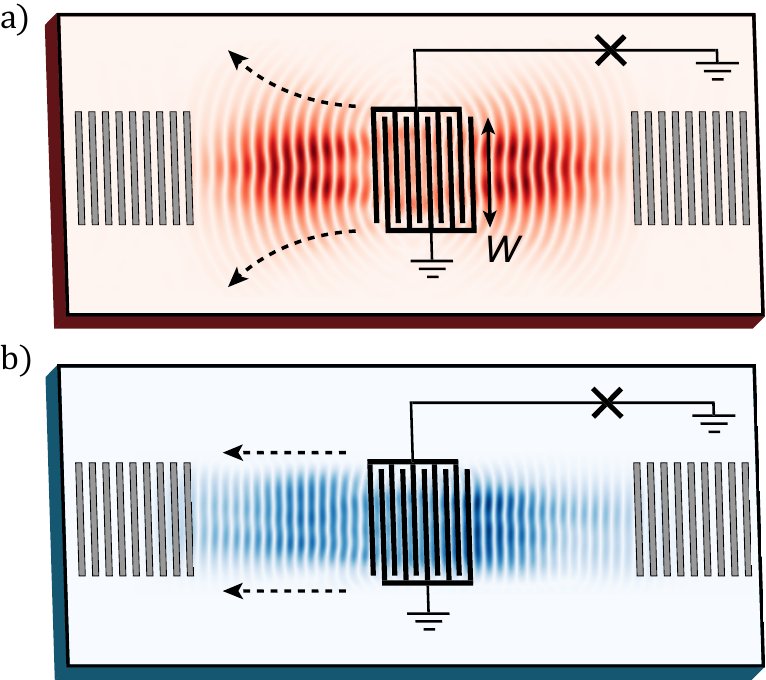}
\caption{Comparison of SAW diffraction on two orientations of quartz. (a) An IDT with aperture $W$ is galvanically connected to a nonlinear circuit element ($\times$) on ST quartz, a common orientation of quartz for SAW devices\cite{morgan2010surface}. A voltage across the IDT launches SAWs (simulated displacement amplitude shown in red) which are confined by Bragg reflectors (gray) to form a multimode acoustic cavity. The IDT, circuit elements and Bragg reflectors are cartoons overlying SAWs shown shortly after launch in time-domain finite-element simulations. As the SAWs propagate they diffract outwards and introduce loss in the resonator. (b) An identical geometry on a minimally-diffracting (MD) orientation of quartz. SAWs (blue) on an MD orientation of quartz diffract less as they propagate, facilitating high mechanical quality factors with small IDT capacitances.} 
\label{fig:Fig1} 
\end{figure}

In a broad array of hybrid acoustic systems, maximizing coupling between a qubit and mechanical degrees of freedom is achieved by tightly confining mechanical strain\cite{maity2020coherent,raniwala2022spin,wang2020coupling,chu2018creation, decrescent2022tightly, heinrich2021quantum}. It can be difficult, however, to confine the strain without significantly increasing mechanical dissipation rates. In the particular example of cQAD with SAW resonators, this increase in dissipation results mainly from surface wave diffraction\cite{sletten2019resolving}. Consider a SAW resonator which is coupled to a nonlinear circuit element via the piezoelectric interaction of an interdigitated transducer (IDT). In such a configuration, shown in Fig. \ref{fig:Fig1}(a), the added IDT capacitance reduces the nonlinearity of the superconducting circuit. Thus for many applications the IDT capacitance must be small, corresponding to a narrow acoustic aperture $W$ on the order of several acoustic wavelengths, $\lambda$. However, acoustic diffraction losses in SAW resonators scale quadratically with the inverse of aperture width causing resonators in this regime to suffer from high loss rates\cite{sletten2019resolving}. This loss significantly limits the mechanical coherence time of hybrid systems based on SAW resonators and thus restricts their capabilities for quantum information processing. 

\begin{figure*}
\centering
\includegraphics[width=\textwidth]{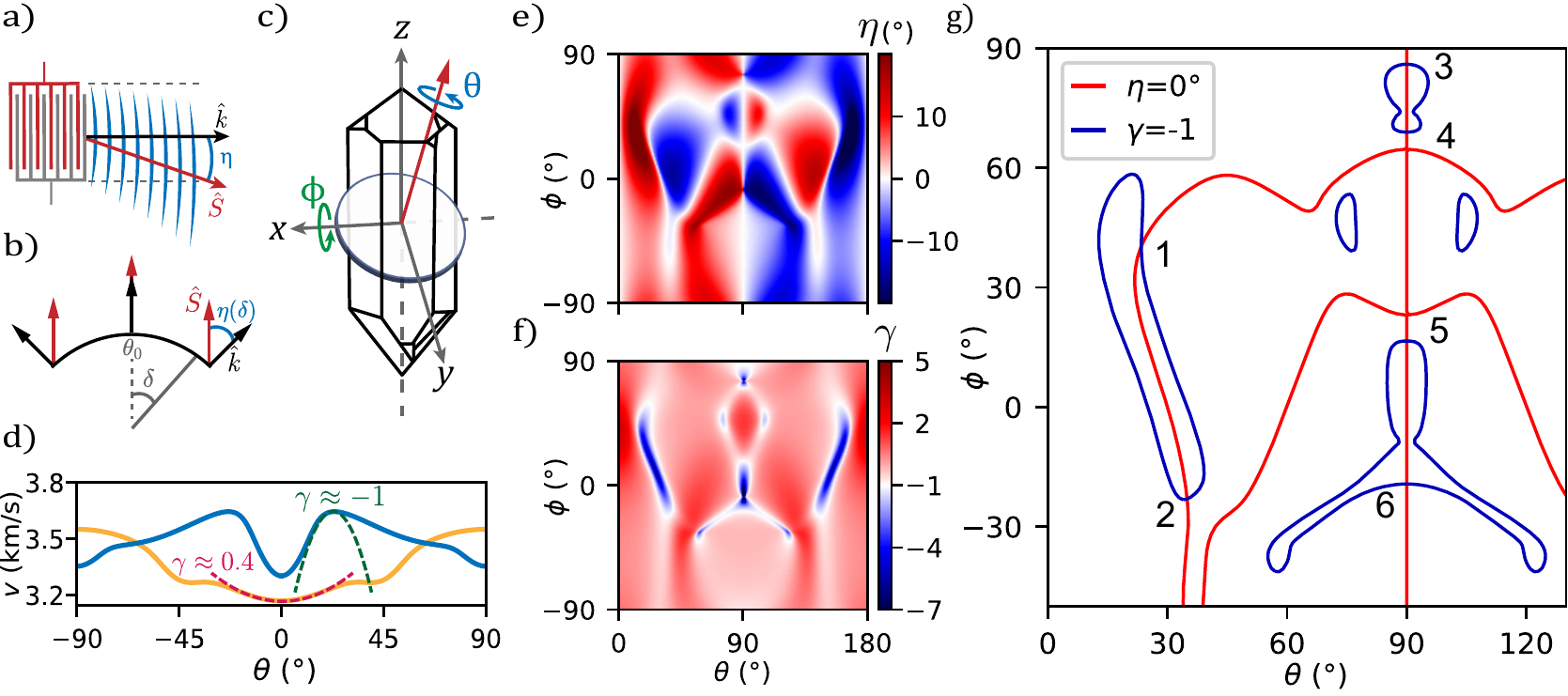}
\caption{Simulated SAW characteristics (a) An IDT launches SAWs on an anisotropic substrate which exhibits beam-steering. (b) Minimal diffraction occurs when $\eta(\delta)=-\delta$ so that $\hat{S}(\delta)$ is constant and transduced SAWs propagate with minimal diffraction spreading. (c) A wafer oriented relative to quartz crystallographic axes: $\phi$ determines the angle at which the wafer is cut from a monocrystalline bar, and $\theta$ determines the planar rotation of a device about the wafer normal. (d) SAW velocity is shown as a function of planar rotation for MD (blue) and ST (orange) quartz. (e) The beam-steering angle $\eta$ is calculated from the simulated velocity profiles for all $\theta$ and $\phi$. (f) From the simulated beam-steering, we calculate the diffraction parameter $\gamma$. (g) The contours for minimal beam-steering (blue) and diffraction (red) are overlaid. The six unique intersections of the curves represent potential MD orientations.}
\label{fig:Fig2}
\end{figure*}

Although it is possible to mitigate the deleterious effects of acoustic diffraction by appropriately curving the cavity boundaries, accomplishing this with piezoelectric media is complicated by their inherent anisotropy\cite{de2003focusing,msall2020focusing,kharel2018ultra}; anisotropy of phase velocity, electromechanical coupling, and reflectivity must all be considered. Rather than compensating for this anisotropy, it is possible to instead exploit it into naturally suppressing acoustic diffraction. At special orientations of some piezoelectric materials, as shown in Fig. \ref{fig:Fig1}(b), the anisotropy will induce an angle-dependent beam-steering such that waves will propagate predominantly along one axis of the substrate. The resulting wavefronts are flat and propagate with minimal diffraction spreading. Substrates which exhibit this property are known as minimally-diffracting (MD) materials\cite{morgan2010surface}. 
 
 Our goal is to find an MD orientation of quartz which is suitable for cQAD platforms. We choose quartz for its extremely low bulk mechanical losses at mK temperatures \cite{goryachev2013observation} and its relatively strong piezoelectricity. An MD orientation of quartz for room-temperature SAW devices was previously identified \cite{Sawtek2000,cowperthwaite2003optimal}, however, the temperature dependence of the piezoelectric and elastic coefficients of quartz causes this cut to lose its MD property as it is cooled to ultra-low temperatures. A new orientation of quartz is required for minimizing SAW diffraction in this regime.

In this work, we use finite element method (FEM) simulations to model diffraction and beam-steering for ultra-low temperature quartz to search for an MD orientation suitable for quantum experiments. From these simulations, we identify an orientation of quartz at Euler angles\cite{institute1978ieee} ($\psi, \phi, \theta$) = ($0^{\circ}, 40.2^{\circ}, 23.4^{\circ}$) which minimizes SAW diffraction and beam-steering to second-order when cooled to ultra-low temperatures. We fabricate SAW resonators with two flat reflectors (see Fig. \ref{fig:Fig1}) on quartz at this orientation which demonstrate high internal quality factors ($\sim$28,000) with very narrow (10$\lambda$) acoustic apertures. This represents a 25$\times$ improvement over equivalent resonators on ST quartz. Slightly wider resonators on this substrate ($W>25\lambda$) demonstrate extremely high internal quality factors ($Q>110,000$); achieving similar quality factors with narrow aperture resonators on non-MD substrates is impossible without complicated anisotropic focusing procedures. We conclude that this orientation, which we name `Cryogenically Optimized Low Diffraction' (COLD) quartz, exhibits minimal diffraction at mK temperatures.

\begin{figure*} 
\centering
\includegraphics[width=\textwidth]{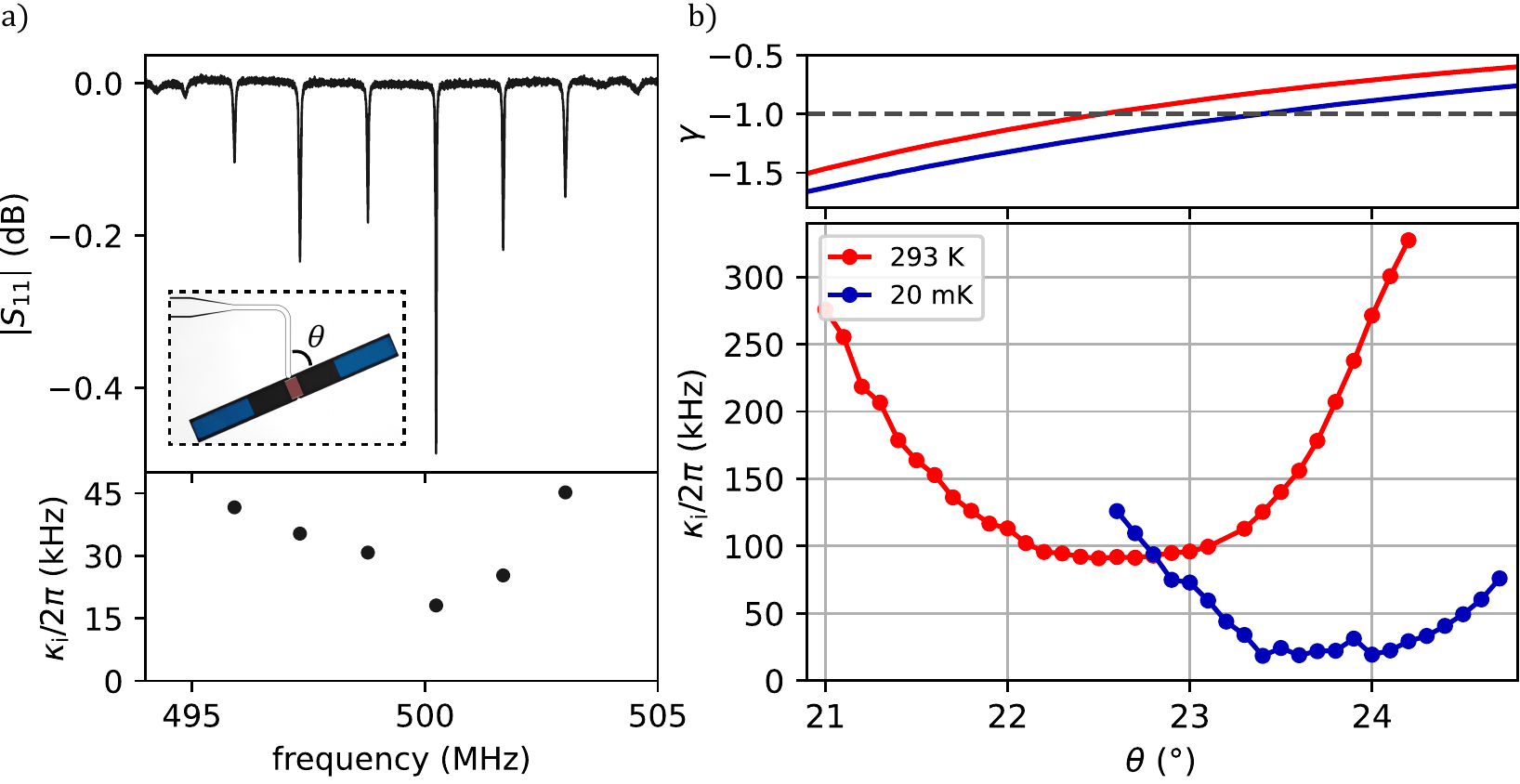}
\caption{Resonator loss and planar rotation. (a) A reflection measurement from a one-port SAW resonator fabricated at $\phi,\theta$ = (40.2$^\circ$,23.4$^\circ$) and mounted on the base plate of a dilution refrigerator cooled to 20 mK. The resonator (false-colored optical image in inset) consists of a double-finger IDT (red) and Bragg mirror gratings (blue) which are designed with a narrow acoustic aperture ($10\lambda$). Aluminium (white) is etched away (black) to define the coplanar waveguide, IDT, and SAW cavity. Internal linewidth for each mode is plotted below. (b) Top: simulated $\gamma$ for SAWs on $\phi=40.2^\circ$ quartz at 293 K (red) and 5 K (blue). Minimal diffraction is predicted at $\theta=22.5^\circ$ (23.4$^\circ$) for warm (cold) devices. Bottom: measured internal linewidth of the central resonator mode of each device as $\theta$ is incremented across many devices. Fit errorbars for all data points are smaller than the marker size. 
}
\label{fig:Fig3}
\end{figure*}

Minimal diffraction in anisotropic materials can be understood as an effect of beam-steering, a phenomenon that causes a beam in an anisotropic medium to propagate in a direction which is not normal to the wavefront. The beam-steering angle $\eta$ is the difference between the directions of the SAW wave vector $\hat{k}$ and energy flow vector $\hat{S}$ (Fig. \ref{fig:Fig2}a). It can be calculated from $v(\theta)$, the SAW phase velocity as a function of planar rotation\cite{morgan2010surface}:
\begin{equation}\label{eq:PFA}
    \eta(\theta) = \mathrm{tan}^{-1}\left(\frac{1}{v(\theta)}\frac{dv(\theta)}{d\theta}\right).
\end{equation}
Beam-steering vanishes at angle $\theta_0$ for which $v(\theta)$ exhibits a stationary point. However, any finite-width IDT oriented at $\theta_0$ will transduce SAWs at a spectrum of angles around $\theta_0$ which, on an anisotropic substrate, necessarily exhibit beam-steering. The second-order diffractive spread of the beam width due to this off-axis beam-steering is characterized by the diffraction parameter $\gamma$,
\begin{equation}\label{eq:DIF}
    \gamma = \frac{d\eta}{d\theta},
\end{equation}
which can be used to calculate the diffraction-limited quality factor $Q_d$ of a flat SAW cavity on an anisotropic substrate\cite{aref2016quantum},
\begin{equation} \label{eq:qDiff}
    Q_d = \frac{5\pi}{|1+\gamma|}(W/\lambda)^2.
\end{equation}
Diffraction is reduced (compared to isotropic substrates) for materials where $-2<\gamma<0$ and otherwise increased\cite{morgan2010surface}. ST quartz, for example, exhibits accelerated wave diffraction $\gamma=0.378$ and isotropic materials exhibit $\gamma=0$. Minimal diffraction occurs when $\gamma=-1$, a special condition such that beam-steering at small angles $\delta$ around some orientation $\theta_0$ follows $\eta(\delta)\approx-\delta$. In this circumstance waves transduced at angle $\theta_0+\delta$ by a finite-width IDT will propagate in the direction $\theta_0$. Thus when $\gamma=-1$, as shown in Fig. \ref{fig:Fig2}(b), $\hat{S}$ points in the same direction irrespective of $\hat{k}$, and SAWs will propagate with minimal diffraction spreading.

We thus search for an orientation of quartz which (i) exhibits $\eta\approx0^\circ$, (ii) exhibits $\gamma\approx-1$, (iii) and is tolerant to potential errors in manufacturer cutting. Due to the trigonal structure of alpha quartz, three Euler angles ($\psi, \phi, \theta$) are necessary to uniquely describe the orientation of a device relative to some crystallographic axes\cite{institute1978ieee}. In this work, however, we consider only cuts where $\psi=0^{\circ}$; this constraint excludes `doubly-rotated' orientations which are more difficult to cut and may result in greater variation of the final substrate orientation. The remaining angles ($\phi, \theta$) are illustrated in Fig. \ref{fig:Fig2}(c): $\phi$ describes the angle between the wafer-normal and the crystallographic Z-axis while $\theta$ corresponds to a planar rotation about the wafer-normal.

Using FEM simulations\cite{COMSOL} we generate velocity curves $v(\phi, \theta)$ in the space ($\psi, \phi, \theta)=(0^{\circ}, -90^{\circ}:90^{\circ}, 0^{\circ}:180^{\circ}$) and apply Eqs. (\ref{eq:PFA}) and (\ref{eq:DIF}) to calculate $\eta$ and $\gamma$. The basis of the simulations is a 3D unit cell with two pairs of periodic mechanical and electric boundary conditions in the longitudinal and transver
se dimensions. The unit cell is one wavelength (7.28 $\mathrm{\mu}$m) wide, one wavelength deep, and three wavelengths tall with an absorbing boundary condition on the bottom surface to mitigate reflection of evanescent SAW energy. Voltage is excited across infinitely-thin terminals to produce a piezoelectric response in the cell without added mass damping, and the terminals use a double-finger structure\cite{aref2016quantum} to mitigate the effects of internal reflections at the fundamental frequency. An eigenfrequency study reveals surface wave modes for a particular orientation, and the velocity of the lowest frequency mode is traced to produce $v(\phi,\theta)$. The crystallographic orientation is specified by rotating the piezoelectric and elastic tensors relative to the unit cell axes. To model low temperature quartz, we use piezoelectric and elastic coefficients for right-handed alpha quartz measured at 5 K\cite{tarumi2007}. Simulated velocity curves for MD ($\phi=40.2^\circ$) and ST ($\phi=-47.25^\circ$) quartz are shown in Fig. \ref{fig:Fig2}(d). Note that these simulations trace only the lowest-order surface wave eigenmodes as we are uninterested in higher-order leaky waves \cite{morgan2010surface}. However, the lowest-order mode is not guaranteed to be a Rayleigh (R) polarized SAW with displacement confined to the sagittal plane; it is possible to have selected a shear-horizontal (SH) or other type SAW. We must thus explicitly check through simulation if an orientation supports R-polarized SAWs.

In Figs. \ref{fig:Fig2}(e) and \ref{fig:Fig2}(f), we plot the values of $\eta$ and $\gamma$ calculated from the simulated velocities. The contours along which beam-steering and diffraction are minimized are plotted in Fig \ref{fig:Fig2}(g). Although the six unique intersections of these curves represent potentially desirable cuts, we find that only one is suitable for further study; the relevant properties for each cut are listed in Table \ref{tab:table1}. To verify that a small error in the wafer cutting will not destroy the MD behavior, we also calculate the sensitivity parameters $\lvert d\eta/d\phi\rvert$ and $\lvert d\gamma/d\phi\rvert$. We find that orientation 1 is Rayleigh-polarized, demonstrates high tolerance to cut errors, exhibits a large electromechanical coupling $k^2$ (Appendix \ref{apdx:electromechanical}), and minimizes diffraction. We name this orientation `COLD' quartz.

\begin{table}
\caption{\label{tab:table1} $\psi=0^\circ$ 5K quartz cuts}
\begin{ruledtabular}
\begin{tabular}{crccccc}
  & $\phi, \theta$ $(^\circ$) & $\lvert d\eta/d\phi\rvert$ & $\lvert d\gamma/d\phi\rvert (^{\circ^{-1}})$&$v$ (m/s) & type & $k^2(\%)$\\
\hline
1 & 40.2, 23.4 & 0.41 & 0.003 & 3643 & R & 0.16\\
2 & -23.1, 34.9 & 0.09 & 0.08 & 3374 & SH & 0.15\\
3 & 86.0, 90.0 & $\sim$ 0 & 0.07 & 3890 & R & $\sim$0\\
4 & 68.9, 90.0 & $\sim$ 0 & 0.37 & 3778 & SH & 0.03\\
5 & 16.5, 90.0 & $\sim$ 0 & 0.17 & 3379 & SH & 0.10\\
6 & -19.3, 90.0 & $\sim$ 0 & 0.2 & 3765 & R & $\sim$0\\
\end{tabular}
\end{ruledtabular}
\end{table}

We procured COLD quartz wafers from a vendor with quoted tolerance ($\pm 0.1^\circ$, $\pm 0.03^\circ$,$\pm 0.1^\circ$). To test the performance of the substrate we first fabricate a series of one-port resonators and vary the angle $\theta$ at which each device is oriented relative to the crystallographic axes. Double-finger IDTs\cite{morgan2010surface} are patterned from a 25 nm aluminum film, and SAWs are confined by etched Bragg mirror gratings with 250 elements and a reflectivity of $1-2\%$ per element. The resonators are designed with wavelength $\lambda=7.28$ $\mathrm{\mu}$m and a mirror-to-mirror cavity length of 150$\lambda$. These resonators are designed with a narrow ($10\lambda$) aperture so that losses are dominated by diffraction. In Fig. \ref{fig:Fig3}(a) we show the spectrum from a microwave reflection measurement of a single $\theta=23.4^\circ$ device mounted on the base plate of a dilution refrigerator which is at temperature of 20 mK. The central mode occurs at 500.24 MHz, corresponding to a 3641.7 m/s speed of sound which is 0.02\% lower than the predicted value of 3642.6 m/s. This mode exhibits an internal linewidth $\kappa_i/2\pi$ = 18.1 kHz. Using Eq. (\ref{eq:qDiff}), this represents a factor of 25 decrease in linewidth compared to an equivalent resonator on ST quartz.

In Fig. \ref{fig:Fig3}(b) we observe a strong dependence of the central mode internal linewidth on $\theta$ as the resonators are rotated in $0.1^\circ$ increments about the wafer surface normal. At room temperature, internal loss is minimized at $\theta=22.5^\circ$ which is in strong agreement with our simulations. As the resonators are cooled, the central angle of minimal loss shifts to $23.7^\circ$. This is higher than the predicted angle of minimal diffraction (23.4$^\circ$), but this discrepancy can be understood as a result of the third-order velocity profile asymmetry. As the aperture width is reduced, higher angle wave vectors with non-symmetric beam-steering increasingly contribute to the transduced wavefront. This produces an additional width-dependent \textit{effective} beam-steering $\eta_\mathrm{eff}$ which is not described by Eq. (\ref{eq:PFA}). This phenomenon shifts the angle of minimal loss from $23.4^\circ$ to $23.7^\circ$, an effect discussed in detail in Appendix \ref{apdx:effBeamSteer}. For wider aperture devices, we expect the angle of minimal loss to shift back to 23.4$^\circ$ as $\eta_\mathrm{eff}$ approaches $0^\circ$.

\begin{figure} 
\centering
\includegraphics[width=\columnwidth]{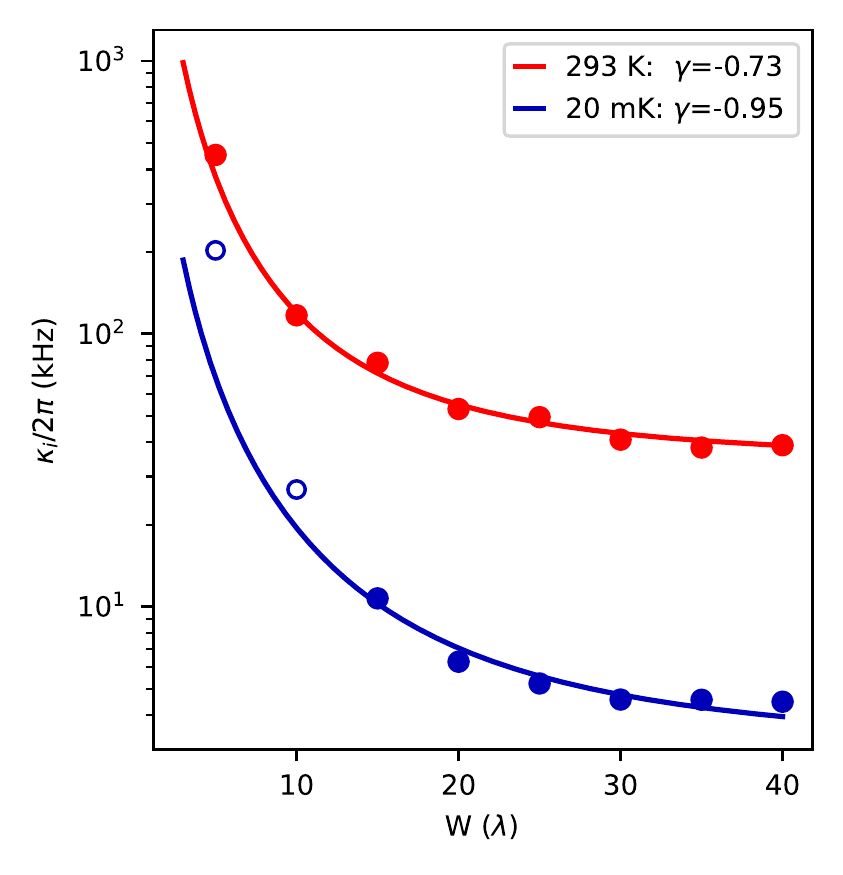}
\caption{Measurement of diffraction parameter. Internal linewidth of SAW resonators measured at room temperature (red) and 20 mK (blue) which sweep the acoustic aperture width. Solid lines show fits to Eq. (\ref{eq:qDiff}), for which we find $\gamma$ = -0.73 at room temperature and $\gamma = -0.95$ at 20 mK. Open circles show widths for which third-order diffraction becomes comparable to the residiual second-order diffraction and Eq. (\ref{eq:qDiff}) becomes incapable of accounting for the loss. Fit errorbars are smaller than the marker size.
}
\label{fig:Fig4}
\end{figure}
To measure $\gamma$, we next fabricate a series of SAW resonators at $\theta$ = $23.4^\circ$ which sweep aperture width from 5$\lambda$ to 40$\lambda$. In Fig. \ref{fig:Fig4}, we observe at both room temperature and 20 mK a monotonic decrease in internal loss as the acoustic aperture of these devices is increased. We fit the data to Eq. (\ref{eq:qDiff}) and find that the substrate exhibits $\gamma$ = $-0.73\pm0.01$ and $\gamma$ = $-0.95\pm0.01$ at room temperature and 20 mK, respectively. At 20 mK, the apparent decrease in loss with increasing aperture width saturates at $W$ = 30$\lambda$, for which the device demonstrates $\kappa_i/2\pi$ = 4.5 kHz. This corresponds to an internal quality factor of 110,000 which is on the order of state-of-the-art SAW resonators. Notably, the fit diverges for the narrowest resonators with $W \lesssim$ 10$\lambda$ measured at 20 mK. At these widths, third-order contributions to diffraction loss from the asymmetric surface velocity begin to dominate and Eq. (\ref{eq:qDiff}) becomes inadequate for describing the losses. 

There are two approaches to reducing the third-order beam-steering losses: compensate for $\eta_\mathrm{eff}$ with device geometry or choose an orientation for which the third derivative of $v(\theta)$ vanishes. Preliminary experiments suggest that it is possible to mitigate $\eta_\mathrm{eff}$ loss by fabricating resonators which have a mirror aperture slightly wider than the IDT aperture to capture the misaligned portion of the beam, however further study on this subject is required. Alternatively, the phenomenon of narrow-aperture-induced effective beam-steering can be eliminated entirely by choosing an orientation which lies along a crystalline symmetry axis so that the velocity profile is fully symmetric about the propagation axis. Unfortunately, of the two symmetry-axis R-polarized SAW orientations identified in Table \ref{tab:table1}, both exhibit extremely low $k^2$. Future work may explore doubly-rotated cuts where $\psi\neq0^\circ$ to search for an MD orientation of quartz which exhibits less 3rd-order asymmetry in $v(\theta)$ and a nonzero $k^2$.

Improving the lifetimes of mechanical resonators coupled to superconducting circuits is critical for maximizing the capabilities of cQAD systems. By discovering a substrate which naturally suppresses SAW diffraction losses, we have provided a solution to the dominant loss source observed in previous SAW-qubit devices \cite{sletten2019resolving}. This will allow this class of devices to be pushed further into the strong-dispersive regime\cite{schuster2007resolving, moores2018cavity, sletten2019resolving} to facilitate the exploration of multimodal quantum information processing. COLD quartz may also be useful in the creation of low-loss cryogenic delay lines\cite{magnusson2015surface, ekstrom2017surface},  phononic waveguides \cite{he2008guiding, fu2019phononic}, and electro-acoustic phase modulators\cite{shao2022electrical}. Future work may search for doubly-rotated orientations of quartz which exhibit greater velocity profile symmetry to suppress the third-order losses suspected to limit the devices studied here. Crystallographic optimization has become less common in recent decades as standard crystal orientations for acoustic devices have emerged. In studying a non-standard orientation of quartz, we highlight the remaining utility of crystallographic engineering for quantum applications.

We thank William Hanson and Hoffman Materials for useful conversations related to this work. We acknowledge support from the Office of Secretary of Defense via the Vannevar Bush Faculty Fellowship, award number N00014-20-1-2833, and the National Science Foundation Physics Frontier Center under grant number PHYS 1734006.

\appendix

\section{Electromechanical coupling}\label{apdx:electromechanical}
\begin{figure} 
\centering
\includegraphics[width=\columnwidth]{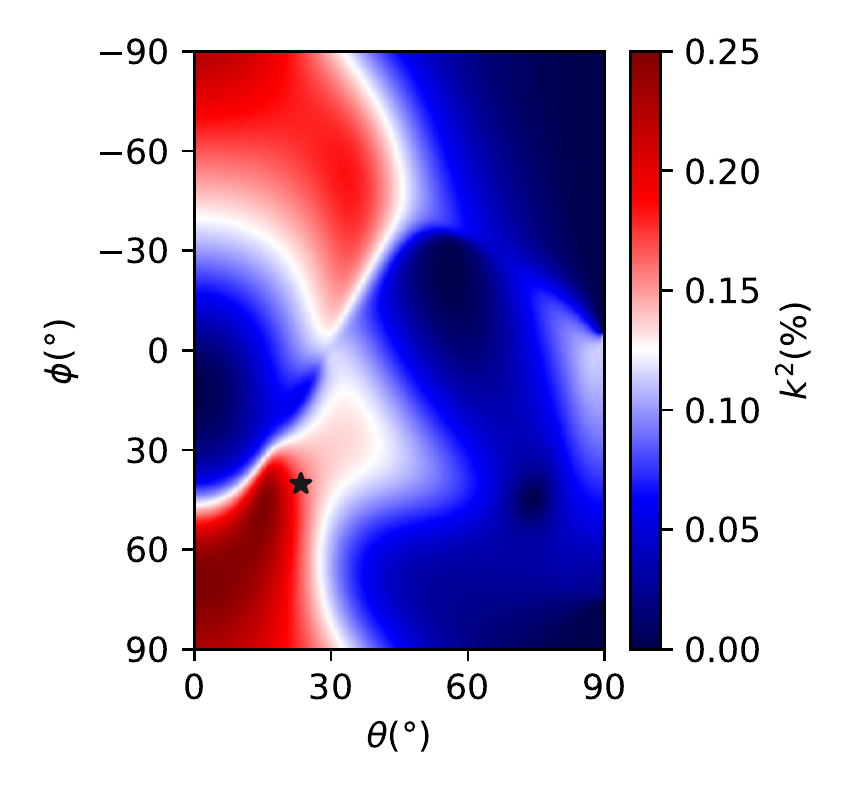}
\caption{Simulated electromechanical coupling coefficient $k^2$ in the $\psi=0^\circ$ parameter space. We simulate COLD quartz (star) to have $k^2=0.161\%$.}
\label{fig:ApdxElectromech}
\end{figure}

The electromechanical coupling coefficient $k^2$ is defined by
\begin{equation} \label{eq:k2}
    k^2 = 2\frac{v_f-v_s}{v_f}
\end{equation}
where $v_f$ and $v_s$ are the free and shorted velocities of a SAW propagating on a substrate\cite{morgan2010surface}. The shorted velocity corresponds to a surface with an idealized metal coating that shorts the longitudinal component of the electric field. To simulate free velocity we remove the electrical terminals from the unit cell and to simulate shorted velocities we add a grounding boundary condition over the entire surface. We use piezoelectric tensors corresponding to room temperature values and repeat the eigenfrequency search with $1^\circ$ resolution over the parameter space.

In Fig. (\ref{fig:ApdxElectromech}), we show the simulated values of $k^2$ for the quartz parameter space where $\psi=0^\circ$. For ST quartz $(0^\circ, -47.25^\circ, 0^\circ)$ we simulate $k^2=0.139\%$, which agrees with the literature value 0.14\%\cite{aref2016quantum}. We simulate that COLD quartz exhibits $k^2=0.161\%$.

\section{Efffective beam-steering}\label{apdx:effBeamSteer}

\begin{figure} 
\centering
\includegraphics[width=\columnwidth]{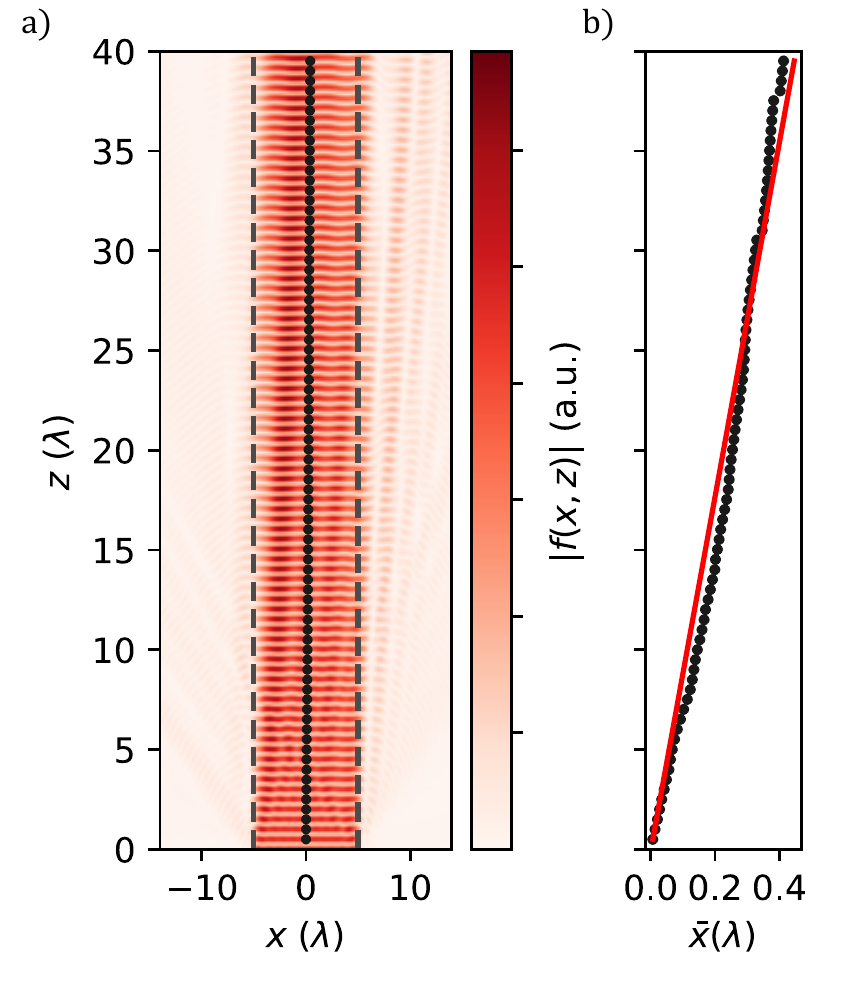}
\caption{Simulated effective beam-steering. (a) The simulated diffractive field for a single $W=10\lambda$ transducer along $z=0$ at orientation $(0^\circ, 40.2^\circ, 23.4^\circ)$. Dashed lines show fictitious cavity boundaries and black circles mark the center of beam intensity $\bar{x}$ for each wavefront. (b) Re-plotted at an appropriate scale, it can be seen that $\bar{x}$ shifts roughly linearly with longitudinal distance $z$. This is fit (red line) to extract $\eta_\mathrm{eff}=0.6^\circ$.}
\label{fig:ApdxFresenel}
\end{figure}
To understand the effect of velocity profile asymmetry on SAWs, we simulate far-field SAW propagation with the angular spectrum of waves technique\cite{kharusi1971diffraction}. The SAW displacement field is proprtional to
\begin{equation}
\label{eq:Fresnel}
    f(x,z) = \frac{1}{\pi}\int^{\infty}_{-\infty}\frac{1}{k_x}\mathrm{sin}(\frac{k_xW}{2})\mathrm{exp}[ik_xx+ik_z(k_x)x]dk_x
\end{equation}
where $x$ and $z$ are the transverse and longitudinal coordinates, $W$ is the width of a transducer launching SAWs along the $x$ axis from $-W/2$ to $W/2$, and $k_x$ and $k_z$ are the components of the wave vector in the $x$ and $z$ directions given that 
\begin{equation}
    k_z(k_x)^2=k(\hat{k})^2-k_x^2,
\end{equation}
where
\begin{equation}
    k(\hat{k})=2\pi f/v(\hat{k}),
\end{equation}
$v(\hat{k})$ is the phase velocity in wave-vector direction $\hat{k}$ and $f$ is the frequency. In Fig. (\ref{fig:ApdxFresenel}), we simulate the diffractive field from a 10$\lambda$ transducer at orientation $(0^\circ, 40.2^\circ, 23.4^\circ$) with a velocity profile generated from FEM unit cell simulations. For each peak in the longitudinal profile along $z$ of the diffractive field, we calculate the weighted center of intensity $\bar{x}$,
\begin{equation}
    \bar{x} = \frac{\int x f(x)^2dx}{\int f(x)^2dx},
\end{equation}
and plot this both on the figure and below as a function of $z$. The center of intensity shifts approximately linearly against $z$, corresponding to an \textit{effective} beam-steering $\eta_\mathrm{eff}\approx0.6^\circ$. 

This phenomenon can be attributed to the asymmetry of the velocity profile. The ideal minimally-diffracting profile $v(\theta)\propto\mathrm{cos}(\theta)$ for all $\theta$ produces an angular beam-steering profile $\eta(\theta)=-\theta$ which facilitates uniaxial transduction of SAWs. However, it is generally sufficient to match the ideal profile for only several degrees about the maximum; wavevector contributions are most significant from these angles. Indeed, the COLD quartz velocity profile approximates a cosine well for $\pm2^\circ$ around $23.4^\circ$ before deviating. This deviation is necessarily asymmetric as COLD quartz is not oriented along a crystalline symmetry axis. In particular, the high-angle side ($\theta>23.4^\circ)$ diverges more quickly and SAWs transduced at these angles do not exhibit large enough beam-steering to achieve an MD wavefront. Contributions from these angles will thus `steer' the propagating SAW towards higher angles, and $\eta_\mathrm{eff}$ will be minimized at some higher angle.
The degree to which a given wave vector-direction $\hat{k}$ contributes to the propagating field is dependent on the width of transducer; this is captured in Eq. (\ref{eq:Fresnel}) by the sinc-term $\frac{1}{k_x}\mathrm{sin}(\frac{k_xW}{2})$. The narrower the IDT aperture is, the more impactful these high-angle deviations become, contributing to a higher angle at which $\eta_\mathrm{eff}$ is minimized. 

\begin{figure} 
\centering
\includegraphics[width=\columnwidth]{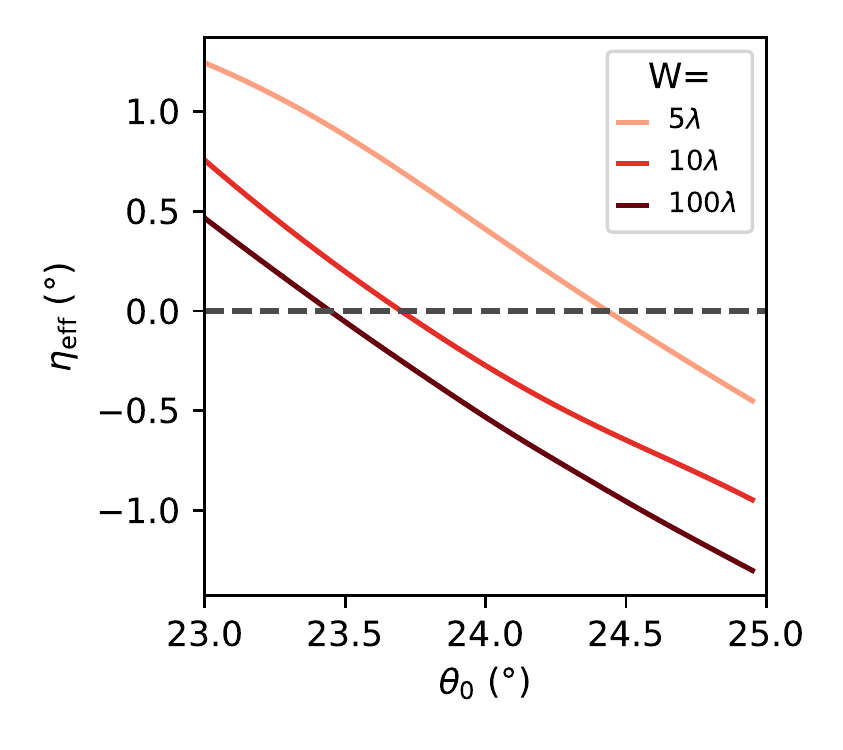}
\caption{Effective beam-steering as a function of IDT orientation $\theta_0$ for transducers of several widths. For crystal orientations with asymmetric velocity profiles, beam-steering becomes a function of width. 
}
\label{fig:ApdxEffEtaWidth}
\end{figure}

In Fig. (\ref{fig:ApdxEffEtaWidth}), we plot the effective beam-steering angle simulated as a function of transduction angle $\theta_0$ for IDTs of several widths. In the case of a very wide aperture ($100\lambda$), the effective beam-steering is minimized at $23.4^\circ$; this transducer is predominantly sampling wave-vectors in the idealized region of the velocity profile and thus agrees with the predictions of Eq. (\ref{eq:PFA}). The $10\lambda$ IDT minimizes effective beam-steering at $23.7^\circ$, which stands in strong agreement with the data shown in Fig. \ref{fig:Fig3}. Reducing the aperture further to $5\lambda$ shifts the angle of minimal effective loss to $24.4^\circ$.

For wider aperture resonators ($W\gtrsim 15\lambda$) this phenomenon is largely negligible. For narrow aperture devices, however, mitigating this added source of loss is necessary. This can be accomplished by shifting the orientation of the transducer to the angle of minimal $\eta_\mathrm{eff}$ or by widening the mirror width compared to the IDT aperture.

\section{Fabrication}\label{apdx:fabrication}
Photoresist patterning is performed with an ASML 5500/100D Wafer Stepper, which supports a 400 nm resolution and 90 nm alignment between lithographic layers. This allows for excellent angular precision while using optical lithography; he smallest device feature size is 910 nm. Wafer preparation begins with a Nanostrip etch to remove residual organics followed by evaporation of a 25 nm aluminum film covering the entire 3" wafer which will later form the metallized features. nLOF 2020 photoresist is then spun onto the wafer, and alignment mark patterns for the wafer stepper are exposed in negative. 10 nm of titanium and 100 nm of gold are evaporated and the films are lifted off to leave behind the alignment marks.

Device fabrication is a simple two layer process for one-port SAW resonators. First, SPR660 photoresist is spun onto the wafer and the stepper is used to pattern regions where the aluminum film will be etched away. This layer defines the CPW, IDT, and cavity regions. The metal is etched with Transene etchant type A. Photoresist is re-spun, and Bragg mirror gratings are patterned in the cavity region. The mirror gratings are etched into the quartz with an Oxford PlasmaLab ICP-380 using flourine. The target depth for this etch was 73 nm or $h/\lambda=1\%$, which was found to yield a $1.5\%$ reflectivity per grating element.

\bibliography{biblio.bib}

\end{document}